\documentclass[a4paper,11pt]{article}
\usepackage[latin1]{inputenc}
\usepackage{indentfirst}
\usepackage[left]{lineno}
\usepackage{amsfonts}
\usepackage{float}

\usepackage{latexsym,amssymb,amsmath}
\usepackage[dvips]{graphics,graphicx,epsfig,color}
\usepackage[lofdepth,lotdepth]{subfig}
\usepackage[lofdepth,lotdepth]{subfig}
\begin{document}
%\linenumbers
\title{\bf A Conservation law for a  sedimentary basin submitted to a seismic wave}
\author{Armand Wirgin\thanks{LMA, CNRS, UPR 7051, Aix-Marseille Univ, Centrale Marseille, F-13453 Marseille Cedex 13, France, ({\tt wirgin@lma.cnrs-mrs.fr})} }
\date{\today}
\maketitle
\begin{abstract}
We establish  the abstract and semi-abstract equations that translate the conservation  law for a totally-filled, or totally-empty basin of arbitrary shape submitted to a SH plane seismic body wave. This law states that the input flux is equal to the sum of the scattered and absorbed fluxes, the latter being equal to zero when the filler medium is non-lossy. We show that the well-known, exact, solutions for the case of a semi-circular basin indeed satisfy this law. The latter can, and should, be employed to test the validity of any theoretical or numerical solution  one might propose for a basin of arbitrary shape.
\end{abstract}
\newline
\newline
Abbreviated title: Conservation of flux relative to the seismic response of a sedimentary basin
\newline
\newline
Corresponding author: Armand Wirgin, \\ e-mail: wirgin@lma.cnrs-mrs.fr
%%%%%%%%%%%%%%%%%%%%%%%%%%%%%%%%%%%%%%%%%%%%%%%%%%%%%%%%%%%%%%%%%%%%%%%%%%%%%%%%%%%%
\newpage
%%%%%%%%%%%%%%%%%%%%%%%%%%%%%%%%%%%%%%%%%%%%%%%%%%%%%%%%%%%%%%%%%%%%%%%%%%%%%%%%%%%%
\tableofcontents
%%%%%%%%%%%%%%%%%%%%%%%%%%%%%%%%%%%%%%%%%%%%%%%%%%%%%%%%%%%%%%%%%%%%%%%%%%%%%%%%%%%%
\newpage
\newpage
%%%%%%%%%%%%%%%%%%%%%%%%%%%%%%%%%%%%%%%%%%%%%%%%%%%%%%%%%%%%%%%%%%%%%%%%%%%%%%%%%%%%%%%%%%%%%%%%%%%%%%
%%%%%%%%%%%%%%%%%%%%%%%%%%%%%%%%%%%%%%%%%%%%%%%%%%%%%%%%%%%%%%%%%%%%%%%%%%%%%%%%%%%%%%%%%%%%%%%%%%%%%%
\section{Introduction}\label{intro}
 Seismic waves  are known to damage  buildings  or industrial facilities on the ground underlain by a below-ground structure (BGS) termed a  sedimentary basin (the latter being filled with a medium that is soft as compared  to the hard medium filling the rest of the below-ground half-space). For this reason it has been (and continues to be) important for theoretical and applied geophysicists to understand the mechanisms of the  seismic response in  and in the vicinity of BGS's. These same researchers have, of course, proposed a great variety of solutions to their specific BGS seismic response problem, but the question that is often ignored or eluded is: how good are these solutions?

 If the underlying boundary-value problem (BVP)is correctly formulated then the ultimate test of whether a solution is valid or not is whether it satisfies the equations inherent to the BVP, but to carry out this test is often very painstaking because it requires generating the field and its gradient at all points of the ground and on the lower, curved, boundary of the basin, not to speak of the field everywhere within and outside of the basin. Another,  manner to test the solution is by finding out if it satisfies a conservation law (such as that of energy). Since we shall be concerned with frequency domain formulations of the BVP, the conservation of energy law takes the form of what we term the conservation of flux law which states that the input flux equals the scattered flux plus the absorbed flux.

 We show that, to employ this law in its first, abstract form, requires computing the scattering amplitude  in the far-field zone (whether the basin filler medium is lossy or non-lossy) as well as the field at all points within the basin (only when the basin is lossy). The latter task (for lossy fillers) is likewise painstaking, so that it is opportune to show (as is done herein) that the said task can be replaced by computing the field and its gradient only on the upper, flat (ground-level) boundary of the basin. To show that the so-contrived conservation (of flux) law makes sense, we apply it to the case of a cylindrical basin of semi-circular shape submitted to a shear-horizontal plane body wave.

 The result of this operation is that the said exact solution indeed satisfies the conservation law and suggests that if this law is not satisfied the solution cannot be qualified as exact (i.e., might even be fallacious if the difference between the input and output (scattered plus absorbed) fluxes is large.

 Thus, what is proposed herein is a method for testing, in rather economical manner, the validity of a solution to a certain class of seismic response problems.
%%%%%%%%%%%%%%%%%%%%%%%%%%%%%%%%%%%%%%%%%%%%%%%%%%%%%%%%%%%%%%%%%%%%%%%%%%%%%%%%%%%%%%%%%%%%%%%%%%%%%%
\section{Description of the seismic scattering problem}\label{desc}
We shall  be concerned  with the flat-ground approximation of the earth's surface and wish to describe  the seismic response to a plane body wave in the case the below-ground region is divided into two parts: 1) a sedimentary basin (assumed to be of infinite extent along one of the cartesian coordinates)  of arbitrary shape,  this meaning that its lower boundary is a curved surface  of arbitrary shape,  and upper boundary a finite portion of the flat ground, and 2) the remainder of the lower half space termed the rock-like underground.

The earthquake sources are assumed to be located in the lower half-space and to be infinitely-distant from the ground so that the seismic (pulse-like) solicitation takes the form of a body (plane) wave in the neighborhood of the basin. This plane wavefield is assumed to be of the shear-horizontal ($SH$) variety, which means that: only one (i.e., the $z$) component of the incident displacement field is non-nil and this field does not depend on $z$.

We shall assume that the basin boundaries do not depend on $z$ and that the (relatively-soft) medium filling the basin as well as the (relatively-hard) medium outside the basin are both linear, homogeneous and isotropic.  It ensues that the scattered and total displacement fields within and outside the basin do not depend on $z$. Thus, the problem we are faced with is 2D ($z$ being the ignorable coordinate), and it is sufficient to search for the $z$-component of the scattered displacement field, designated by $u_{z}^{s}(\mathbf{x};\omega)$ in the sagittal (i.e., $x-y$) plane, when $u_{z}^{i}(\mathbf{x};\omega)$ designates the incident displacement field, with $\mathbf{x}=(x,y)$ and $\omega=2\pi f$  the angular frequency, $f$ the frequency. The temporal version of the displacement field is $u_{z}(\mathbf{x};t)=2\Re\int_{0^{\infty}}u_{z}^{i}(\mathbf{x};\omega)\exp(-i\omega t)d\omega$ wherein $t$ is the temporal variable.
\begin{figure}[ht]
\begin{center}
\includegraphics[width=0.75\textwidth]{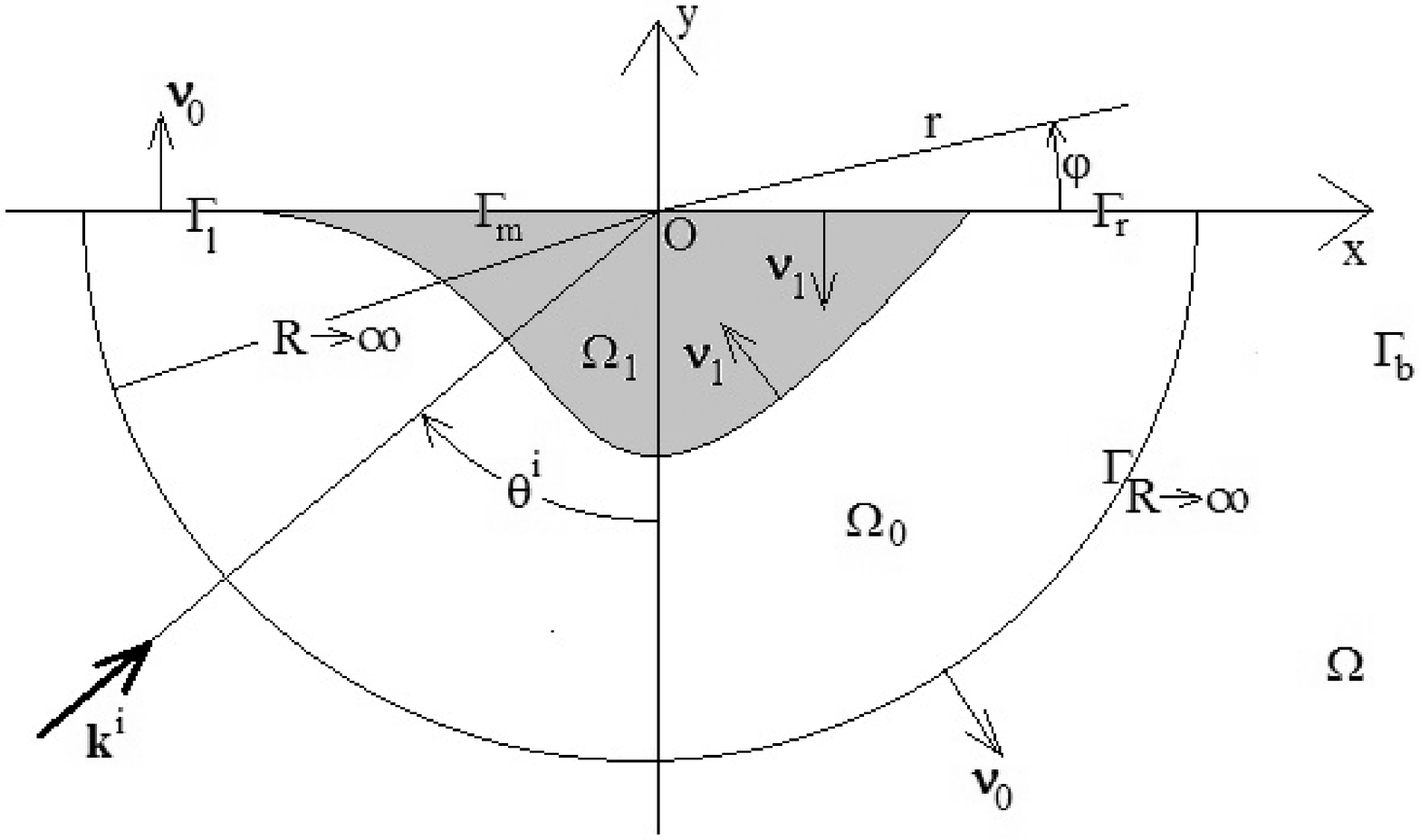}
\caption{Sagittal plane view of 2D scattering configuration. }
\label{basin}
\end{center}
\end{figure}

Fig. \ref{basin} describes the scattering configuration in the sagittal plane. In this figure, $\mathbf{k}^{i}=\mathbf{k}^{i}(\theta^{i},\omega)$ is the incident wavevector oriented so that its $z$  component is nil, and $\theta^{i}$ is the angle of incidence.

The totality of the ground is stress-free but since the basin is assumed to be in welded contact with the surrounding below-ground medium, its boundary is the locus of continuous displacement and stress, this meaning that the incident field penetrates within the basin as well as being scattered outside the basin in the remaining lower half space.

 This study does not impose any restrictions either on the shape of the lower boundary or on the (lossy or non-lossy) nature  of the basin.

The two media (other than the one of the upper half space, being occupied by the vacuum, is of no interest since the field cannot penetrate therein)  are $M^{[l]}~;l=0,1$ within which the real shear moduli $\mu^{[l]}$  and the generally-complex shear- body wave velocities are $\beta^{[l]}~;~l=0,1$  i.e., $\beta^{[l]}=\beta^{'[l]}+i\beta^{''[l]}$, with $\beta^{'[l]}\ge 0$, $\beta^{''[l]}\le 0$, $\beta^{[l]}=\sqrt{\frac{\mu^{[l]}}{\rho^{[l]}}}$, and  $\rho^{[l]}$ the (generally-complex) mass density. The shear-wave velocity $\beta^{[0]}$ is assumed to be real, i.e., $\beta^{''[0]}=0$.
%%%%%%%%%%%%%%%%%%%%%%%%%%%%%%%%%%%%%%%%%%%%%%%%%%%%%%%%%%%%%%%%%%%%%%%%%%%%%%%%%%%%%%%%%%%%%%%%%%%%%%
\section{The boundary  value problem}\label{bvp}
The basin occupies (in the sagittal plane (SP)) the finite-sized region $\Omega_{1}$.  The remainder of the below-ground half-space occupies the region $\Omega_{0}$. $\Omega_{0}$ is entirely filled with $M^{[0]}$ and $\Omega_{1}$ is entirely filled with $M^{[1]}$.

Always in the sagittal plane, the flat ground is designated  by $\Gamma_{g}$ (described by $y=0$, with $x,y$ the cartesian coordinates in the SP) and is composed  of three segments; $\Gamma_{l}$, $\Gamma_{m}$, and $\Gamma_{r}$, which designate the left-hand, middle, and right-hand portions respectively of $\Gamma_{g}$. The basin is a below-ground structure whose upper and lower boundaries (in the SP) are $\Gamma_{m}$ and $\Gamma_{b}$.

The analysis takes place in the space-frequency framework, so that all constitutive and field variables depend on the frequency $f$. This dependence will henceforth be implicit (e.g., $u_{z}(\mathbf{x},f)$, with $\mathbf{x}=(x,y)$, will be denoted by $u(\mathbf{x})$).

The seismic solicitation is an incident shear-horizontal (SH)  plane wave field of the form
\begin{equation}\label{1-000}
u^{i}(\mathbf{x})=a^{i}\exp(i\mathbf{k}^{i}\cdot\mathbf{x})=a^{i}\exp[i(k_{x}^{i}x+k_{z}^{i}y)]~,
\end{equation}
wherein $a^{i}=a^{i}(\omega)$ is the spectral amplitude of the seismic pulse, $\mathbf{k}^{i}=(k_{x}^{i},k_{y}^{i})$, $k_{x}^{i}=k^{[0]}\sin\theta^{i}$, $k_{y}^{i}=k^{[0]}\cos\theta^{i}$, $k^{[l]}=\omega/\beta^{[l]}~;~l=0,1,2$.

Owing to the fact that the configuration comprises two distinct regions, each in which the elastic parameters are constants as a function of the space variables, it is opportune to employ domain decomposition and (later on) separation of variables. Thus, we decompose the total  field $u$ as:
\begin{equation}\label{1-010}
u(\mathbf{x})=u^{[l]}(\mathbf{x})~;~\forall\mathbf{x}\in\Omega_{l},~l=0,1~,
\end{equation}
with the understanding that these fields satisfy the 2D SH frequency domain elastic wave equation (i.e., Helmholtz equation)
\begin{equation}\label{1-020}
\Big(\triangle+\big(k^{[l]}\big)^{2}\Big)u^{[l]}(\mathbf{x})=0~;~\forall\mathbf{x}\in\Omega_{l},~l=0,1~,
\end{equation}
 with the notations $\triangle=\frac{\partial^{2}}{\partial x^{2}}+\frac{\partial^{2}}{\partial y^{2}}$ in the cartesian coordinate system of the sagittal plane.

 In addition, the field $u^{[0]}$ satisfies the radiation condition
\begin{equation}\label{1-030}
u^{[0]}(\mathbf{x})-u^{i}(\mathbf{x})\sim \text {outgoing~wave} ~;~\|\mathbf{x}\|\rightarrow \infty~.
\end{equation}
due to the fact that $\Omega_{0}$ is unbounded (i.e., a semi-infinite domain).

The stress-free nature of the boundaries $\Gamma_{l}$, $\Gamma_{m}$, $\Gamma_{r}$, entail the boundary conditions:
\begin{equation}\label{bc-010}
\mu^{[0]}u_{,y}^{[0]}(\mathbf{x})=0~;~\forall\mathbf{x}\in\Gamma_{l}+\Gamma_{r}~,
\end{equation}
\begin{equation}\label{bc-020}
\mu^{[1]}u_{,y}^{[2]}(\mathbf{x})=0~;~\forall\mathbf{x}\in\Gamma_{m}~,
\end{equation}
wherein   $u_{,\zeta}$ denotes the first  partial derivative of $u$ with respect to $\zeta$.

Finally, the fact that $\Gamma_{b}$ was assumed to be an interface, across which two media are in welded contact, entails the continuity conditions:
\begin{equation}\label{bc-040}
u^{[0]}(\mathbf{x})-u^{[1]}(\mathbf{x})=0~;~\forall\mathbf{x}\in\Gamma_{b}~,
\end{equation}
\begin{equation}\label{bc-050}
\mu^{[0]}u_{,y}^{[0]}(\mathbf{x})-\mu^{[1]}u_{,y}^{[1]}(\mathbf{x})=0~;~\forall\mathbf{x}\in\Gamma_{b}~,
\end{equation}

The purpose of addressing such a boundary-value problem is to determine $u^{[l]}(\mathbf{x});~l=0,1$ for various solicitations  and parameters  relative to the  geometries of, and media filling, $\Omega_{l}~;~l=0,1$. The analysis which follows is less-ambitious in that  its purpose is only to establish a conservation law governing $u^{[l]}(\mathbf{x})~;~l=0,1$ without actually solving for the latter. However, to show that this conservation law makes sense, we shall appeal to the well-known rigorous, closed-form solution of the problem in which the basin is of semi-circular shape.
%%%%%%%%%%%%%%%%%%%%%%%%%%%%%%%%%%%%%%%%%%%%%%%%%%%%%%%%%%%%%%%%%%%%%%%%%%%%%%%%%%%%%%%%%%%%%%%%%%%%%%
\section{Basic ingredients of the conservation of flux law}\label{bicl}
Eq. (\ref{1-020}) yields
\begin{equation}\label{2-020}
\Big(\triangle+\big[\big(k^{[l]}\big)^{2}\big]^{*}\Big)u^{[l]*}(\mathbf{x})=0~;~\forall\mathbf{x}\in\Omega_{l},~l=0,1~,
\end{equation}
wherein $(X+iY)^{*}=X-iY$. It follows (with $d\Omega$ the differential surface element) that
\begin{equation}\label{2-030}
\int_{\Omega_{l}}u^{[l]*}\mathbf{x})\Big\{\Big(\triangle+\big(k^{[l]}\big)^{2}\Big)u^{[l]}(\mathbf{x})-
u^{[l]*}(\mathbf{x})\Big(\triangle+\big[\big(k^{[l]}\big)^{2}\big]^{*}\Big)u^{[l]}(\mathbf{x})\Big\}d\Omega
=0~;~\forall\mathbf{x}\in\Omega_{l},~l=0,1~,
\end{equation}
or
\begin{multline}\label{2-040}
\int_{\Omega_{l}}\Big\{u^{[l]*}(\mathbf{x})\triangle u^{[l]}(\mathbf{x})-u^{[l]}(\mathbf{x})\triangle u^{*[l]}(\mathbf{x})\Big\}d\Omega+\\
\int_{\Omega_{l}}\Big\{\big(k^{[l]}\big)^{2}-\big[\big(k^{[l]}\big)^{2}\big]^{*}\Big\}\|u^{[l]}(\mathbf{x})\|^{2}d\Omega
=0~;~\forall\mathbf{x}\in\Omega_{l},~l=0,1~.
\end{multline}
We want to apply Green's second identity to the first integral, and to do this we must define the (closed) boundaries of $\Omega_{l}$. We already know that $\Omega_{1}$, constituted by the union of $\Gamma_{m}$ and $\Gamma_{b}$, is closed, but until now, $\Omega_{0}$ was not closed. To close it, we imagine a semicircle $\Gamma_{\mathcal{R}}$, of large radius $\mathcal{R}$ (taken to be infinitely large in the limit), centered at the origin $O$,  to be drawn so as to intersect the ground at $x=\pm \mathcal{R}$ and to intersect the $y$ axis at $y=-\mathcal{R}$. Thus, the closed boundaries of $\Omega_{1}$ are:
\begin{equation}\label{2-050}
\partial_{\Omega_{1}}=\Gamma_{m}\cup\Gamma_{b}~~,~~
\partial_{\Omega_{0}}=\Gamma_{l}\cup\Gamma_{b}\cup\Gamma_{r}\cup\Gamma_{\mathcal{R}\rightarrow\infty}~,
\end{equation}
and we shall designate by $\boldsymbol{\nu}_{l}$ the unit vector normal to $\partial\Omega_{l}$  that points towards the exterior of $\partial\Omega_{l}$. We now apply Green's second identity to obtain (with $d\Gamma$ the differential arc element)
\begin{multline}\label{2-060}
\int_{\partial\Omega_{l}}\Big\{u^{[l]*}(\mathbf{x})\boldsymbol{\nu}_{l}\cdot\nabla u^{[l]}(\mathbf{x})-u^{[l]}(\mathbf{x}) \boldsymbol{\nu}_{l}\cdot\nabla u^{*[l]}(\mathbf{x})\Big\}d\Omega+\\
\int_{\Omega_{l}}\Big\{\big(k^{[l]}\big)^{2}-\big[\big(k^{[l]}\big)^{2}\big]^{*}\Big\}\|u^{[l]}(\mathbf{x})\|^{2}d\Omega
=0~;~\forall\mathbf{x}\in\Omega_{l},~l=0,1~.
\end{multline}
or, in condensed form (on account of the non-lossy nature of the solid filling $\Omega_{0}$ and the homogeneous nature of the solid filling $\Omega_{1}$):
\begin{equation}\label{2-070}
\Im\int_{\partial\Omega_{0}}u^{[0]*}(\mathbf{x})\boldsymbol{\nu}_{0}\cdot\nabla u^{[0]}(\mathbf{x})d\Gamma=0~,
\end{equation}
\begin{equation}\label{2-080}
\Im\int_{\partial\Omega_{1}}u^{[1]*}(\mathbf{x})\boldsymbol{\nu}_{1}\cdot\nabla u^{[1]}(\mathbf{x})d\Gamma-\Im\big[\big(k^{[1]}\big)^{2}\big]\int_{\Omega_{1}}\|u^{[1]}(\mathbf{x})\|^{2}d\Omega=0~.
\end{equation}
More explicitly, and on account of (\ref{2-050}),
\begin{multline}\label{2-090}
\Im\int_{\Gamma_{l}+\Gamma_{r}}u^{[0]*}(\mathbf{x})\boldsymbol{\nu}_{0}\cdot\nabla u^{[0]}(\mathbf{x})d\Gamma+
\Im\int_{\Gamma_{b}}u^{[0]*}(\mathbf{x})\boldsymbol{\nu}_{0}\cdot\nabla u^{[0]}(\mathbf{x})d\Gamma+\\
\Im\int_{\Gamma_{\infty}}u^{[0]*}(\mathbf{x})\boldsymbol{\nu}_{0}\cdot\nabla u^{[0]}(\mathbf{x})d\Gamma=0~,
\end{multline}
\begin{equation}\label{2-100}
\Im\int_{\Gamma_{m}}u^{[1]*}(\mathbf{x})\boldsymbol{\nu}_{1}\cdot\nabla u^{[1]}(\mathbf{x})d\Gamma+
\Im\int_{\Gamma_{b}}u^{[1]*}(\mathbf{x})\boldsymbol{\nu}_{1}\cdot\nabla u^{[1]}(\mathbf{x})d\Gamma-\Im\big[\big(k^{[1]}\big)^{2}\big]\int_{\Omega_{1}}\|u^{[1]}(\mathbf{x})\|^{2}d\Omega=0~.
\end{equation}
Due to  $\boldsymbol{\nu}_{0}(\mathbf{x})=-\boldsymbol{\nu}_{1}(\mathbf{x})~;~\forall\mathbf{x}\in\Gamma_{b}$,
and the boundary and continuity conditions, we have:
\begin{equation}\label{2-110}
u^{[0]*}(\mathbf{x})\boldsymbol{\nu}_{0}\cdot\nabla u^{[0]}(\mathbf{x})=
-\frac{\mu^{[1]}}{\mu^{[0]}}u^{[1]*}(\mathbf{x})\boldsymbol{\nu}_{0}\cdot\nabla u^{[1]}(\mathbf{x})~;~\forall\mathbf{x}\in\Gamma_{b}~,
\end{equation}
\begin{equation}\label{2-115}
u^{[0]*}(\mathbf{x})\boldsymbol{\nu}_{0}\cdot\nabla u^{[0]}(\mathbf{x})=0~;~\forall\mathbf{x}\in\Gamma_{l}+\Gamma_{r}~,
\end{equation}
\begin{equation}\label{2-120}
u^{[1]*}(\mathbf{x})\boldsymbol{\nu}_{1}\cdot\nabla u^{[1]}(\mathbf{x})=0~;~\forall\mathbf{x}\in\Gamma_{m}~,
\end{equation}
so that, since $\frac{\mu^{[1]}}{\mu^{[0]}}$
was assumed to be real,  (\ref{2-090})-(\ref{2-100}) become:
\begin{equation}\label{2-130}
-\Im\int_{\Gamma_{b}}u^{[1]*}(\mathbf{x})\boldsymbol{\nu}_{1}\cdot\nabla u^{[1]}(\mathbf{x})d\Gamma+
\frac{\mu^{[0]}}{\mu^{[1]}}\Im\int_{\Gamma_{\infty}}u^{[0]*}(\mathbf{x})\boldsymbol{\nu}_{0}\cdot\nabla u^{[0]}(\mathbf{x})d\Gamma=0~,
\end{equation}
\begin{equation}\label{2-140}
\Im\int_{\Gamma_{b}}u^{[1]*}(\mathbf{x})\boldsymbol{\nu}_{1}\cdot\nabla u^{[1]}(\mathbf{x})d\Gamma-
\Im\big[\big(k^{[1]}\big)^{2}\big]\int_{\Omega_{1}}\|u^{[1]}(\mathbf{x})\|^{2}d\Omega=0~.
\end{equation}
The sum of these two equations yields
\begin{equation}\label{2-150}
\Im\int_{\Gamma_{\infty}}u^{[0]*}(\mathbf{x})\boldsymbol{\nu}_{0}\cdot\nabla u^{[0]}(\mathbf{x})d\Gamma-
\frac{\mu^{[1]}}{\mu^{[0]}}\Im\big[\big(k^{[1]}\big)^{2}\big]\int_{\Omega_{1}}\|u^{[1]}(\mathbf{x})\|^{2}d\Omega=0~,
\end{equation}
or, on account of (\ref{2-140}),
\begin{equation}\label{2-160}
\Im\int_{\Gamma_{\infty}}u^{[0]*}(\mathbf{x})\boldsymbol{\nu}_{0}\cdot\nabla u^{[0]}(\mathbf{x})d\Gamma
-\frac{\mu^{[1]}}{\mu^{[0]}}\Im\int_{\Gamma_{b}}u^{[1]*}(\mathbf{x})\boldsymbol{\nu}_{1}\cdot\nabla u^{[1]}(\mathbf{x})d\Gamma=0~.
\end{equation}
Eqs. (\ref{2-150})-(\ref{2-160}) are two alternate expressions of the same sought-for conservation law.
%%%%%%%%%%%%%%%%%%%%%%%%%%%%%%%%%%%%%%%%%%%%%%%%%%%%%%%%%%%%%%%%%%%%%%%%%%%%%%%%%%%%%%%%%%%%%%%%%%%%%%%%%%%%%%%%%%%%%%%%%%%%%%%%%%%%%%%
\section{Asymptotic far-field expression of the field in $\Omega_{0}$}
The fact that the conservation law involves the field on $\Gamma_{\infty}=\Gamma_{\mathcal{R}\rightarrow\infty}$ means that we must seek an expression for the far-field $\tilde{u}^{[0]}(\mathbf{x})=u^{[0]}(\mathbf{x})~;~\mathcal{R}\rightarrow\infty$ which at least satisfies the Helmholtz equation, the radiation condition and the stress-free boundary condition on $\Gamma_{l}$ and $\Gamma_{r}$. To do this, we employ separation of variables (SOV) in the cylindrical coordinate system.

In this $(r,\phi,z)$ system, with $z$ the ignorable coordinate:
\begin{equation}\label{3-010}
\boldsymbol{\nu}_{0}\big|_{\Gamma_{l}+\Gamma_{r}}=\mathbf{e}_{r}~~,~~\nabla u=\mathbf{e}_{r}u_{,r}+\mathbf{e}_{\phi}r^{-1}u_{,\phi}~~,~~\triangle u=u_{,rr}+r^{-1}u_{,r}+r^{-2}u_{,\phi\phi}~,
\end{equation}
so that the Helmholtz equation becomes
\begin{equation}\label{3-020}
u^{[0]}_{,rr}+r^{-1}u^{[0]}_{,r}+r^{-2}u^{[0]}_{,\phi\phi}+\big(k^{[0]}\big)^{2}u^{[0]}=0~,
\end{equation}
and the stress-free boundary condition takes the form
\begin{equation}\label{3-030}
\mu^{[0]}\boldsymbol{\nu}_{0}\cdot\nabla u\big|_{\Gamma_{l}+\Gamma_{r}}=0~\Rightarrow~
\mu^{[0]}u_{,\phi}\big|_{\phi=0}=\mu^{[0]}u_{,\phi}\big|_{\phi=\pi}=0~.
\end{equation}
Suppose that $\Gamma_{b}$ can be described by the equation $r=\gamma(\phi)~;~\forall\phi\in[\pi,2\pi]$. Then, the SOV representation of the field in the region $\Omega_{0}^{+}$, defined by $r\ge\max_{\phi\in[\pi,2\pi]}\gamma(\phi)$, that satisfies: the Helmholtz equation (\ref{3-020}), the boundary condition (\ref{3-030}) and the radiation condition (\ref{1-030}) is \cite{tr71,wi95b}
\begin{equation}\label{3-040}
u^{[0]}(\mathbf{x})-u^{i}(\mathbf{x})=\sum_{m=0}^{\infty}\mathcal{B}_{m}H_{m}^{(1)}(k^{[0]}r)\cos(m\phi)~;~\forall\mathbf{x}\in\Omega_{0}^{+}~.
\end{equation}
wherein $H_{m}^{(1)}( )$ is the $m$-th order Hankel function of the first kind and the $\mathcal{B}_{m}$ are  yet-undetermined coefficients that: i) are constants with respect to $\mathbf{x}=(r,\phi)$, and ii)  depend only on: the shape and dimensions of $\Gamma_{b}$, the frequency $f$, $a^{i}$, and $\theta^{i}$ (all implicit in (\ref{3-040})).

It is easy to understand that
\begin{equation}\label{3-050}
u^{r}(\mathbf{x})=a^{i}\exp[i(k_{x}x-k_{y}y)]~,
\end{equation}
is also a solution of the Helmholtz equation which satisfies the radiation condition so that we can write (\ref{3-050})) in the alternate form:
\begin{equation}\label{3-060}
u^{[0]}(\mathbf{x})-u^{i}(\mathbf{x})-u^{r}(\mathbf{x})=u^{s}(\mathbf{x})=\sum_{m=0}^{\infty}b_{m}H_{m}^{(1)}(k^{[0]}r)\cos(m\phi)~;~\forall\mathbf{x}\in\Omega_{0}^{+}~.
\end{equation}
wherein the $b_{m}$ have the same properties as the $\mathcal{B}_{m}$ and $u^{s}$ is the so-called scattered field in $\Omega_{0}$ that is composed of a sum of outgoing cylindrical waves.

The key property of each of these waves is their asymptotic behavior for $r\rightarrow\infty$, which is obtained via the well-known formula 9.2.3 in \cite{as68}:
\begin{equation}\label{3-070}
H_{m}^{(1)}(k^{[0]}r)\sim\Big(\frac{2}{\pi k^{[0]}r)}\Big)^{1/2}\exp[i(k^{[0]}r-m\pi/2-\pi/4)]~;~k^{[0]}r\rightarrow\infty~.
\end{equation}
whence
\begin{equation}\label{3-080}
u^{s}(\mathbf{x})\sim=B(\phi)\Big(\frac{2}{\pi k^{[0]}r)}\Big)^{1/2}\exp[i(k^{[0]}r-\pi/4)]~;~k^{[0]}r\rightarrow\infty~.
\end{equation}
wherein
\begin{equation}\label{3-090}
B(\phi)=\sum_{m=0}^{\infty}b_{m}\exp(-im\pi/2)\cos(m\phi)~.
\end{equation}
%
%%%%%%%%%%%%%%%%%%%%%%%%%%%%%%%%%%%%%%%%%%%%%%%%%%%%%%%%%%%%%%%%%%%%%%%%%%%%%%%%%%%%%%%%%%%%%%%%%%%%%%%%%%%%%%%%%%%%%%%%%%%%%%%%%%%%%%%
\section{Detailed form of the conservation law}
Now let us return to (\ref{2-150})-(\ref{2-160}) which both can be written as
\begin{equation}\label{3-100}
I-J=0~.
\end{equation}
with
\begin{equation}\label{3-110}
I=\Im\int_{\Gamma_{\infty}}u^{[0]*}(\mathbf{x})\boldsymbol{\nu}_{0}\cdot\nabla u^{[0]}(\mathbf{x})d\Gamma~,
\end{equation}
and let us examine $I$ more closely. On account of (\ref{3-060}) we have
\begin{equation}\label{3-120}
I=\Im\int_{\Gamma_{\infty}}[u^{*}_{i}+u^{*}_{r}+u^{*}_{s}]\boldsymbol{\nu}_{0}\cdot\nabla [u_{i}+u_{r}+u_{s}]d\Gamma=
(I^{ii}+I^{rr})+(I^{ir}+I^{ri})+(I^{is}+I^{si})+(I^{rs}+I^{sr})+I^{ss}~,
\end{equation}
or, due to the facts that:
                                 $d\Gamma\big|_{\Gamma_{\mathcal{R}}}=\mathcal{R}d\phi$ and
$\boldsymbol{\nu}_{0}\cdot\nabla u^{[0]}\big|_{\Gamma_{\mathcal{R}}}=
u_{,r}^{[0]}(\mathcal{R},\phi)$,
we obtain
\begin{equation}\label{3-130}
I=\Im~\lim_{\mathcal{R}\rightarrow\infty}\int_{\pi}^{2\pi}[u^{i*}(\mathcal{R},\phi)+u^{r*}(\mathcal{R},\phi)+u^{s*}(\mathcal{R},\phi)]
[
u_{,r}^{i}(\mathcal{R},\phi)+
u_{,r}^{r}(\mathcal{R},\phi)+
u_{,r}^{s}(\mathcal{R},\phi)
]
\mathcal{R}d\phi~,
\end{equation}
in which we must adopt the following cylindrical coordinate representations of $u^{i}$ and $u^{r}$, wherein $x=r\cos\phi$ and $y=r\sin\phi$:
\begin{equation}\label{3-140}
  u^{i}=a^{i}\exp[ir(k_{x}^{i}\cos\phi+k_{y}^{i}\sin\phi)]~~,
~~u^{r}=a^{i}\exp[ir(k_{x}^{i}\cos\phi-k_{y}^{i}\sin\phi)]~.
\end{equation}
It is then straightforward to find:
\begin{multline}\label{3-160}
I^{ii}=\Im~\lim_{\mathcal{R}\rightarrow\infty}\int_{\pi}^{2\pi}u^{i*}(\mathcal{R},\phi)u_{,r}^{i}(\mathcal{R},\phi)\mathcal{R}d\phi=
\lim_{\mathcal{R}\rightarrow\infty}\|a^{i}\|^{2}2k^{i}\mathcal{R}~,\\
I^{rr}=\Im~\lim_{\mathcal{R}\rightarrow\infty}
\int_{\pi}^{2\pi}u^{r*}(\mathcal{R},\phi)u_{,r}^{r}(\mathcal{R},\phi)\mathcal{R}d\phi=
\lim_{\mathcal{R}\rightarrow\infty}(-\|a^{i}\|^{2})2k_{x}^{i}\mathcal{R}~,
\end{multline}
whence
\begin{equation}\label{3-170}
I^{ii}+I^{rr}=0~.
\end{equation}
Likewise, we find:
\begin{multline}\label{3-180}
I^{ir}=\Im~\lim_{\mathcal{R}\rightarrow\infty}\int_{\pi}^{2\pi}u^{i*}(\mathcal{R},\phi)u_{,r}^{r}(\mathcal{R},\phi)\mathcal{R}d\phi=\\
\|a^{i}\|^{2}\Im~\lim_{\mathcal{R}\rightarrow\infty}i\mathcal{R}\int_{\pi}^{2\pi}(k_{x}^{i}\cos\phi-k_{y}^{i}\sin\phi)\exp(-2ik_{y}^{i}\mathcal{R}\sin\phi)d\phi~.
\end{multline}
\begin{multline}\label{3-190}
I^{ri}=\Im~\lim_{\mathcal{R}\rightarrow\infty}\int_{\pi}^{2\pi}u^{r*}(\mathcal{R},\phi)u_{,r}^{i}(\mathcal{R},\phi)\mathcal{R}d\phi=\\
\|a^{i}\|^{2}\Im~\lim_{\mathcal{R}\rightarrow\infty}i\mathcal{R}\int_{\pi}^{2\pi}(k_{x}^{i}\cos\phi+k_{y}^{i}\sin\phi)\exp(2ik_{y}^{i}\mathcal{R}\sin\phi)d\phi~.
\end{multline}
which become, with the change of variable $\alpha=\phi-\frac{3\pi}{2}$:
\begin{equation}\label{3-200}
I^{ir}=
\|a^{i}\|^{2}\Im~\lim_{\mathcal{R}\rightarrow\infty}i\mathcal{R}\int_{-\pi/2}^{\pi/2}(k_{x}^{i}\sin\alpha+k_{y}^{i}\cos\alpha)\exp(2ik_{y}^{i}\mathcal{R}\cos\alpha)d\alpha~.
\end{equation}
\begin{equation}\label{3-210}
I^{ri}=
\|a^{i}\|^{2}\Im~\lim_{\mathcal{R}\rightarrow\infty}i\mathcal{R}\int_{-\pi/2}^{\pi/2}(k_{x}^{i}\sin\alpha-k_{y}^{i}\cos\alpha)\exp(-2ik_{y}^{i}\mathcal{R}\cos\alpha)d\alpha~.
\end{equation}
so that
\begin{multline}\label{3-220}
I^{ir}+I^{ri}=
\|a^{i}\|^{2}\Im~\lim_{\mathcal{R}\rightarrow\infty}i\mathcal{R}\int_{-\pi/2}^{\pi/2}2k_{x}^{i}\sin\alpha~\cos(2k_{y}^{i}\mathcal{R}\cos\alpha)d\alpha+\\
\|a^{i}\|^{2}\Im~\lim_{\mathcal{R}\rightarrow\infty}i\mathcal{R}\int_{-\pi/2}^{\pi/2}2ik_{y}^{i}\cos\alpha~\sin(2k_{y}^{i}\mathcal{R}\cos\alpha)d\alpha~,
\end{multline}
in which the first integral vanishes because the integrand is an odd function of $\alpha$ and the second term vanishes because the imaginary part of a real quantity is nil. Thus
\begin{equation}\label{3-230}
I^{ir}+I^{ri}=0~.
\end{equation}
The next step is to consider:
\begin{multline}\label{3-240}
I^{is}=\Im~\lim_{\mathcal{R}\rightarrow\infty}\int_{\pi}^{2\pi}u^{i*}(\mathcal{R},\phi)u_{,r}^{s}(\mathcal{R},\phi)\mathcal{R}d^\phi=
\Im~\lim_{\mathcal{R}\rightarrow\infty}\int_{\pi}^{2\pi}u^{i*}(\mathcal{R},\phi)\tilde{u}_{,r}^{s}(\mathcal{R},\phi)\mathcal{R}d\phi=
\\
\Im~\lim_{\mathcal{R}\rightarrow\infty}\int_{\pi}^{2\pi}(a^{i})^{*}\exp[-iR(k_{x}^{i}\cos\phi+k_{y}^{i}\sin\phi)]
ik^{[0]}B(\phi)\Big(\frac{2}{\pi k^{[0]}\mathcal{R}}\Big)^{1/2}\exp[i(k^{[0]}\mathcal{R}-\pi/4)]\mathcal{R}d\phi=
\\
\Im~(a^{i})^{*}ik^{[0]}\lim_{\mathcal{R}\rightarrow\infty}\mathcal{R}\Big(\frac{2}{\pi k^{[0]}\mathcal{R}}\Big)^{1/2}\exp[i(k^{[0]}\mathcal{R}-\pi/4)\mathcal{I}^{is}~.
\end{multline}
and
\begin{multline}\label{3-250}
I^{rs}=\Im~\lim_{\mathcal{R}\rightarrow\infty}\int_{\pi}^{2\pi}u^{r*}(\mathcal{R},\phi)u_{,r}^{s}(\mathcal{R},\phi)\mathcal{R}d\phi=
\Im~\lim_{\mathcal{R}\rightarrow\infty}\int_{\pi}^{2\pi}\tilde{u}^{r*}(\mathcal{R},\phi)\tilde{u}_{,r}^{s}(\mathcal{R},\phi)\mathcal{R}d\phi=
\\
\Im~\lim_{\mathcal{R}\rightarrow\infty}\int_{\pi}^{2\pi}(a^{i})^{*}\exp[-iR(k_{x}^{i}\cos\phi-k_{y}^{i}\sin\phi)]
ik^{[0]}B(\phi)\Big(\frac{2}{\pi k^{[0]}\mathcal{R}}\Big)^{1/2}\exp[i(k^{[0]}\mathcal{R}-\pi/4)]\mathcal{R}d\phi=
\\
\Im~(a^{i})^{*}ik^{[0]}\lim_{\mathcal{R}\rightarrow\infty}\mathcal{R}\Big(\frac{2}{\pi k^{[0]}\mathcal{R}}\Big)^{1/2}\exp[i(k^{[0]}\mathcal{R}-\pi/4)\mathcal{I}^{rs}~.
\end{multline}
in which
\begin{multline}\label{3-250}
\mathcal{I}^{is}(k^{[0]}r)=\int_{\pi}^{2\pi}B(\phi)\exp[-iR(k_{x}^{i}\cos\phi+k_{y}^{i}\sin\phi)]d\phi~;~k^{[0]}r\rightarrow\infty\\
\mathcal{I}^{rs}(k^{[0]}r)=\int_{\pi}^{2\pi}B(\phi)\exp[-iR(k_{x}^{i}\cos\phi-k_{y}^{i}\sin\phi)]d\phi~;~k^{[0]}r\rightarrow\infty~.
\end{multline}
These integrals, whose integrands contain a large parameter which make the integrands highly oscillatory,  are of the type that can be treated by the method of stationary phase (MSP) \cite{er56}.

Consider the integral
\begin{equation}\label{3-260}
f(\chi)=\int_{a}^{b}g(t)\exp[i\chi h(t)]dt~,
\end{equation}
in which $\chi$ is a large positive variable and $h(t)$  a real function of the real variable $t$. Suppose that $g$ is continuous and $h$ is twice continuously differentiable. If the following conditions are satisfied: \\\\
\begin{equation}\label{3-270}
 \tau \text{~is the only stationary point of $h$}~,
\end{equation}
\begin{equation}\label{3-270}
a<\tau<b~,
\end{equation}
\begin{equation}\label{3-280}
h_{,t}|_{t=\tau}=0~,
\end{equation}
\begin{equation}\label{3-290}
h_{,tt}|_{t=\tau}>0~,
\end{equation}
then the result of the MPS  is:
\begin{equation}\label{3-300}
f(\chi)\sim\left[\frac{2\pi}{\chi h_{,tt}|_{t=\tau}}\right]^{1/2}g(\tau)\exp[i\chi h(\tau)+i\pi/4]~;~\chi\rightarrow\infty~,
\end{equation}
Note that if the condition (\ref{3-290}) is not satisfied, but rather $h_{,tt}|_{t=\tau}<0$,  then it suffices to apply the MPS to $f^{*}$ and then take the complex conjugate of the MPS result.

Now, let us return to $\mathcal{I}^{is}(k^{[0]}\mathcal{R})$, which can be written as
\begin{equation}\label{3-310}
\mathcal{I}^{is}(k^{[0]}\mathcal{R})=\int_{\pi}^{2\pi}B(\phi)\exp[-ik^{[0]}\mathcal{R}\sin(\theta^{i}+\phi)]d\phi~,
\end{equation}
in which; $\chi=k^{[0]}\mathcal{R}$, $g=B(\phi)$, $h=-\sin(\theta^{i}+\phi)$, $\tau=\Phi$, so that $h_{,\phi}=-cos(\theta^{i}+\phi)$ which is nil for $\theta^{i}+\phi=(2n+1)\pi/2; n=0,\pm 1,...$. Also, $h_{,\phi\phi}=\sin((\theta^{i}+\phi)$, which, on account of the fact that $\theta^{i}\in [-\pi/2,\pi/2]$, cannot be positive for any $n$. Thus, instead we try to apply the MSP to
\begin{equation}\label{3-320}
\mathcal{I}^{is*}(k^{[0]}\mathcal{R})=\int_{\pi}^{2\pi}B^{*}(\phi)\exp[ik^{[0]}\mathcal{R}\sin(\theta^{i}+\phi)]d\phi~,
\end{equation}
for which: $\chi=k^{[0]}\mathcal{R}$, $g=B(\phi)$, $h=-in(\theta^{i}+\phi)$, so that $h_{,\phi}=\cos(\theta^{i}+\phi)$ which again vanishes for $\theta^{i}+\phi=(2n+1)\pi/2; n=0,\pm 1,...$.  Also, $h_{,\phi\phi}=-\sin((\theta^{i}+\phi)$, which is positive, and equal to +1,  for  $\Phi=3\pi/2-\theta^{i}$ which also satisfies (\ref{3-270}). Also, $h(\Phi)=1$, so that the MSP gives rise to
\begin{equation}\label{3-330}
\mathcal{I}^{is*}(k^{[0]}\mathcal{R})\sim
\left[\frac{2\pi}{k^{[0]}\mathcal{R}}\right]^{1/2}B^{*}(3\pi/2-\theta^{i})
\exp[-ik^{[0]}\mathcal{R}-i\pi/4)]~;~k^{[0]}\mathcal{R}\rightarrow\infty~,
\end{equation}
whence
\begin{equation}\label{3-340}
\mathcal{I}^{is}(k^{[0]}\mathcal{R})\sim
\left[\frac{2\pi}{k^{[0]}\mathcal{R}}\right]^{1/2}B(3\pi/2-\theta^{i})
\exp[ik^{[0]}\mathcal{R}+i\pi/4)]~;~k^{[0]}\mathcal{R}\rightarrow\infty~,
\end{equation}
so that
\begin{equation}\label{3-350}
I^{is}=2\Re~a^{i*}B(3\pi/2-\theta^{i})\exp(2ik^{[0]}\mathcal{R})
~;~k^{[0]}\mathcal{R}\rightarrow\infty~,
\end{equation}
In the same way we find:
\begin{equation}\label{3-360}
\mathcal{I}^{rs}(k^{[0]}\mathcal{R})\sim
\left[\frac{2\pi}{k^{[0]}\mathcal{R}}\right]^{1/2}B^{*}(3\pi/2+\theta^{i})
\exp[-ik^{[0]}\mathcal{R}+i\pi/4)]~;~k^{[0]}\mathcal{R}\rightarrow\infty~,
\end{equation}
\begin{equation}\label{3-370}
I^{rs}=2\Re~a^{i*}B(3\pi/2+\theta^{i})~,
\end{equation}
The next step is to consider:
\begin{multline}\label{3-380}
I^{si}=\Im~\lim_{\mathcal{R}\rightarrow\infty}\int_{\pi}^{2\pi}u^{s*}(\mathcal{R},\phi)u_{,r}^{i}(\mathcal{R},\phi)\mathcal{R}d\phi=
\Im~\lim_{\mathcal{R}\rightarrow\infty}\int_{\pi}^{2\pi}u^{s*}(\mathcal{R},\phi)\tilde{u}_{,r}^{i}(\mathcal{R},\phi)\mathcal{R}d\phi=
\\
\Im~\lim_{\mathcal{R}\rightarrow\infty}\int_{\pi}^{2\pi}
a^{i}i(k_{x}^{i}\cos\phi+k_{y}^{i}\sin\phi)\exp[i\mathcal{R}(k_{x}^{i}\cos\phi+k_{y}^{i}\sin\phi)]\times\\
B^{*}(\phi)\Big(\frac{2}{\pi k^{[0]}\mathcal{R}}\Big)^{1/2}\exp[-i(k^{[0]}\mathcal{R}-\pi/4)]\mathcal{R}d\phi=
\Im~a^{i}ik^{[0]}\lim_{\mathcal{R}\rightarrow\infty}\mathcal{R}\Big(\frac{2}{\pi k^{[0]}\mathcal{R}}\Big)^{1/2}\exp[-i(k^{[0]}\mathcal{R}-\pi/4)\mathcal{I}^{si}~.
\end{multline}
in which
\begin{equation}\label{3-390}
\mathcal{I}^{si}=\frac{1}{k^{[0]}}\int_{\pi}^{2\pi}B^{*}(\phi)(k_{x}^{i}\cos\phi+k_{y}^{i}\sin\phi)
\exp[i\mathcal{R}(k_{x}^{i}\cos\phi+k_{y}^{i}\sin\phi)]d\phi
~;~k^{[0]}\mathcal{R}\rightarrow\infty~,
\end{equation}
The application of the MSP leads to
\begin{equation}\label{3-400}
\mathcal{I}^{si}\sim
-\left[\frac{2\pi}{k^{[0]}\mathcal{R}}\right]^{1/2}B^{*}(3\pi/2-\theta^{i})
\exp[-ik^{[0]}\mathcal{R}+i\pi/4)]~;~k^{[0]}\mathcal{R}\rightarrow\infty~,~,
\end{equation}
so that
\begin{equation}\label{3-410}
I^{si}=\lim_{\mathcal{R}\rightarrow\infty}
(-2)\Re~a^{i}B^{*}(3\pi/2-\theta^{i})\exp(-2ik^{[0]}\mathcal{R})~.
\end{equation}
It follows that:
\begin{multline}\label{3-420}
I^{is}+I^{si}=\lim_{\mathcal{R}\rightarrow\infty}~2\Re~[a^{i*}B(3\pi/2-\theta^{i})\exp(2ik^{[0]}\mathcal{R})-
a^{i}B^{*}(3\pi/2-\theta^{i})\exp(-2ik^{[0]}\mathcal{R})]=\\
\lim_{\mathcal{R}\rightarrow\infty}~2\Re~\big[2i\Im~[a^{i*}B(3\pi/2-\theta^{i})\sin(2ik^{[0]}\mathcal{R})]\big]~,
\end{multline}
or, on account of the fact that the real part of an imaginary number is nil,
\begin{equation}\label{3-430}
I^{is}+I^{si}=0~.
\end{equation}
The next step is to evaluate
\begin{multline}\label{3-440}
I^{sr}=\Im~\lim_{\mathcal{R}\rightarrow\infty}\int_{\pi}^{2\pi}u^{s*}(\mathcal{R},\phi)u_{,r}^{r}(\mathcal{R},\phi)\mathcal{R}d\phi=
\Im~\lim_{\mathcal{R}\rightarrow\infty}\int_{\pi}^{2\pi}\tilde{u}^{s*}(\mathcal{R},\phi)\tilde{u}_{,r}^{r}
(\mathcal{R},\phi)\mathcal{R}d\phi=
\\
\Im~\lim_{\mathcal{R}\rightarrow\infty}\int_{\pi}^{2\pi}
a^{i}i(k_{x}^{i}\cos\phi-k_{y}^{i}\sin\phi)\exp[iR(k_{x}^{i}\cos\phi-k_{y}^{i}\sin\phi)]\times\\
B^{*}(\phi)\Big(\frac{2}{\pi k^{[0]}\mathcal{R}}\Big)^{1/2}\exp[-i(k^{[0]}\mathcal{R}-\pi/4)]\mathcal{R}d\phi=
\Im~a^{i}ik^{[0]}\lim_{\mathcal{R}\rightarrow\infty}\mathcal{R}\Big(\frac{2}{\pi k^{[0]}\mathcal{R}}\Big)^{1/2}\exp[-i(k^{[0]}\mathcal{R}-\pi/4)\mathcal{I}^{sr}~.
\end{multline}
in which
\begin{multline}\label{3-450}
\mathcal{I}^{sr}(k^{[0]}\mathcal{R})=\frac{1}{k^{[0]}}\int_{\pi}^{2\pi}
B^{*}(\phi)(k_{x}^{i}\cos\phi-k_{y}^{i}\sin\phi)\exp[iR(k_{x}^{i}\cos\phi-k_{y}^{i}\sin\phi)]d\phi~;~k^{[0]}\mathcal{R}\rightarrow\infty~.
\end{multline}
The application of the MSP yields:
\begin{equation}\label{3-460}
\mathcal{I}^{sr}\sim B^{*}(3\pi/2+\theta^{i})\left(\frac{2\pi}{k^{[0]}R}\right)^{1/2}\exp[i(k^{[0]}R-\pi/4)] ~;~k^{[0]}\mathcal{R}\rightarrow\infty~.
\end{equation}
whence
\begin{equation}\label{3-470}
I^{sr}=2\Re~a^{i} B^{*}(3\pi/2+\theta^{i})~,
\end{equation}
which entails
\begin{equation}\label{3-475}
I^{rs}+I^{sr}=2\Re~\big[a^{i*}B(3\pi/2+\theta^{i})~+a^{i}B^{*}(3\pi/2+\theta^{i})\big]=2\Re\big[2\Re~a^{i*}B(3\pi/2+\theta^{i})\big]~,
\end{equation}
or
\begin{equation}\label{3-480}
I^{rs}+I^{sr}=4\Re~\big[a^{i*}B(3\pi/2+\theta^{i})\big]~.
\end{equation}
The next-to-last step is to examine
\begin{multline}\label{3-490}
I^{ss}=\lim_{\mathcal{R}\rightarrow\infty}\Im~\int_{\pi}^{2\pi}
u^{s*}(\mathcal{R},\phi)u_{,\mathcal{R}}^{s}(\mathcal{R},\phi)\mathcal{R}d\phi=\\
\lim_{\mathcal{R}\rightarrow\infty}\Im~\int_{\pi}^{2\pi}
\Big[B^{*}(\phi)\Big(\frac{2}{\pi k^{[0]}\mathcal{R}}\Big)^{1/2}
\exp|-i(k^{[0]}\mathcal{R}-\pi/4)]
\Big]\times\\
\Big[ik^{[0]}B(\phi)\Big(\frac{2}{\pi k^{[0]}\mathcal{R}}\Big)^{1/2}
\exp|i(k^{[0]}\mathcal{R}-\pi/4)]\Big]\mathcal{R}d\phi
~,
\end{multline}
or
\begin{equation}\label{3-495}
I^{ss}=\frac{2}{\pi}\int_{\pi}^{2\pi}\|B(\phi)\|^{2}d\phi~.
\end{equation}
The last step is to carry out the sum in (\ref{3-120}), in which we employ (\ref{3-170}), (\ref{3-230}), (\ref{3-430}), (\ref{3-480}), and (\ref{3-495}), to obtain
\begin{equation}\label{3-500}
I=\frac{2}{\pi}\int_{\pi}^{2\pi}\|B(\phi)\|^{2}d\phi+4\Re~\big[a^{i*}B(3\pi/2+\theta^{i})\big]~,
\end{equation}
which is the detailed expression of the term $I$ in the conservation law (\ref{2-160}), which we wrote as $I-J=0$.

Let us now examine in detail the term $J$ of this law. The definition of $J$ is
\begin{equation}\label{2-150}
J=\frac{\mu^{[1]}}{\mu^{[0]}}\Im\int_{\Gamma_{b}}u^{[1]*}(\mathbf{x})\boldsymbol{\nu}_{1}\cdot\nabla u^{[1]}(\mathbf{x})d\Gamma~,
\end{equation}
but, as underlined earlier, our ambition was not to solve the boundary-value problem, namely for the field $u^{[1]}(\mathbf{x})$ within the basin, so that we cannot go beyond expressing the conservation law as
\begin{equation}\label{3-500}
I-J=\frac{2}{\pi}\int_{\pi}^{2\pi}\|B(\phi\|^{2}d\phi+4\Re~\big[a^{i*}B(3\pi/2+\theta^{i})\big]-
\frac{\mu^{[1]}}{\mu^{[0]}}\Im\int_{\Gamma_{b}}u^{[1]*}(\mathbf{x})\boldsymbol{\nu}_{1}\cdot\nabla u^{[1]}(\mathbf{x})d\Gamma=0~,
\end{equation}
whose meaning, is unfortunately not clear for the moment. To cope with this problem, we make use of the alternative expression (\ref{2-150}) of the conservation law
\begin{equation}\label{2-510}
I-K=\Im\int_{\Gamma_{\infty}}u^{[0]*}(\mathbf{x})\boldsymbol{\nu}_{0}\cdot\nabla u^{[0]}(\mathbf{x})d\Gamma-
\frac{\mu^{[1]}}{\mu^{[0]}}\Im\big[\big(k^{[1]}\big)^{2}\big]\int_{\Omega_{1}}\|u^{[1]}(\mathbf{x})\|^{2}d\Omega=0~,
\end{equation}
which authorizes us to write (\ref{2-150}) as
\begin{equation}\label{3-520}
I-K=\frac{2}{\pi}\int_{\pi}^{2\pi}\|B(\phi)\|^{2}d\phi+4\Re~\big[a^{i*}B(3\pi/2+\theta^{i})\big]-
\frac{\mu^{[1]}}{\mu^{[0]}}\Im\big[\big(k^{[1]}\big)^{2}\big]\int_{\Omega_{1}}\|u^{[1]}(\mathbf{x})\|^{2}d\Omega=0~.
\end{equation}
 This expression informs us that if the medium within the ~~basin ~~domain ~~$\Omega_{1}$ is ~~non-lossy, ~~~then~~~ $\Im k^{[1]}=0$ ~~so~ that ~~~ $K=\frac{\mu^{[1]}}{\mu^{[0]}}\Im\big[\big(k^{[1]}\big)^{2}\big]\int_{\Omega_{1}}\|u^{[1]}(\mathbf{x})\|^{2}d\Omega=0$,~~ which~~ entails
$J=\frac{\mu^{[1]}}{\mu^{[0]}}\Im\int_{\Gamma_{b}}u^{[1]*}(\mathbf{x})\boldsymbol{\nu}_{1}\cdot\nabla u^{[1]}(\mathbf{x})d\Gamma=0$. This means that $J$ accounts for damping (i.e., absorption) in $\Omega_{1}$ since $J$ vanishes in the absence of a loss mechanism, as expressed by $c^{[1]}$ (and therefore $k^{[1]}$ being complex. A valid question is then: damping/absorption of what? It is not energy because the units of $J=K$ are not those of energy, but it is something related to energy, which we shall name 'flux'. If we reason in terms of energy, the only means by which energy can be lost in a scattering problem such as ours is by radiation damping, which is the mechanism by which (scattered) energy escapes to the outer reaches of $\Omega_{0}$, i.e., escapes to $r\rightarrow\infty$ in the half-space $y<0$ due to the fact that this half-space is not bounded for negative $y$. As stated earlier, radiation damping is exclusively related to the scattered wave portion of the total field, and the function that expresses this relation to the scattered field is $B(\phi)$, so that the the term expressing radiation damping must be $\frac{2}{\pi}\int_{\pi}^{2\pi}\|B(\phi)\|^{2}d\phi$ in $I$. We are therefore authorized to call this term the 'scattered flux'. Now to continue to reason in terms of energy, it is pertinent to ask: if 'lost energy',  is the sum of absorbed and radiation damping energies, then what is the 'provided energy', if we accept the fact that energy must be conserved, i.e.,  'lost energy'='provided energy'? In our problem, the means by which energy is provided is obviously via the incident plane wave, or in other terms: the means by which flux is provided is via the incident plane wave and the only term in our 'conservation of flux' relation that can account for this is $4\Re~\big[a^{i*}B(3\pi/2+\theta^{i})\big]$ in $I$. Since this term also contains a quantity related to the scattered field (via $B(3\pi/2+\theta^{i})$ ) it is more coherent to normalize the expression of the conservation of flux law $I-J=0$ by dividing it by $-4\Re~\big[a^{i*}B(3\pi/2+\theta^{i})\big]$ so as to obtain
\begin{equation}\label{3-530}
\frac{\frac{-2}{\pi}\int_{\pi}^{2\pi}\|B(\phi)\|^{2}d\phi}{4\Re~\big[a^{i*}B(3\pi/2+\theta^{i})\big]}+
\frac{\frac{\mu^{[1]}}{\mu^{[0]}}\Im\int_{\Gamma_{b}}u^{[1]*}(\mathbf{x})\boldsymbol{\nu}_{1}\cdot\nabla u^{[1]}(\mathbf{x})d\Gamma}
{4\Re~\big[a^{i*}B(3\pi/2+\theta^{i})\big]}=1~,
\end{equation}
which can be written as
\begin{equation}\label{3-540}
\mathcal{S}+\mathcal{A}=\mathcal{I}~,
\end{equation}
wherein
\begin{equation}\label{3-550}
\mathcal{S}=\frac{-\int_{\pi}^{2\pi}\|B(\phi)\|^{2}\frac{d\phi}{\pi}}
{2\Re~\big[a^{i*}B(3\pi/2+\theta^{i})\big]}=\frac{\mathcal{-J}}{\mathcal{K}}~,
\end{equation}
is the so-called 'normalized scattered flux',
\begin{equation}\label{3-560}
\mathcal{A}=\frac{\frac{\mu^{[1]}}{2\mu^{[0]}}\Im\int_{\Gamma_{b}}u^{[1]*}(\mathbf{x})\boldsymbol{\nu}_{1}\cdot\nabla u^{[1]}(\mathbf{x})d\Gamma}
{2\Re~\big[a^{i*}B(3\pi/2+\theta^{i})\big]}=\frac{\mathcal{L}}{\mathcal{K}}~,
\end{equation}
is the so-called 'normalized absorbed flux', and
\begin{equation}\label{3-570}
\mathcal{I}=1~.
\end{equation}
is the so-called 'normalized incident flux'. In black-box language, we can say the the conservation law expresses the fact that the 'input flux' $\mathcal{I}$ equals the 'output flux' $\mathcal{S}+\mathcal{A}$, the latter being the sum of the scattered flux $\mathcal{S}$ and the 'absorbed flux' $\mathcal{A}$, wherein, for convenience we have dropped the term 'normalized' which is henceforth implicit.

Note that (\ref{3-540})-(\ref{3-570}) is in agreement with the conservation of flux relation previously obtained in \cite{wi73} for the case of a non-lossy basin (i.e., $\mathcal{A}=0$ filled with a medium that is identical to the one occupying $\Omega_{0}$).

Note also that in the case the basin is non-lossy, the conservation law does not depend explicitly on any of the constitutive parameters of the basin, i.e.,  $c^{[1]}=c^{'[1]}$ and $\mu^{[1]}$. However it does depend implicitly on these parameters via $B(\phi)$.
%%%%%%%%%%%%%%%%%%%%%%%%%%%%%%%%%%%%%%%%%%%%%%%%%%%%%%%%%%%%%%%%%%%%%%%%%%%%%%%%%%%%%%%%%%%%%%%%%%%%%%%%%%%%%%%%%%%
\section{Demonstration that the conservation of flux relation is satisfied by the well-known solution of a  basin scattering problem}
The exact, explicit solution of the scattering problem for the semi-circular basin with center at the midpoint of $\Gamma_{m}$, whose shape is described by $r=\gamma(\phi)=a;~\forall\phi[\pi,2\pi]$, with $a$ the  radius, submitted to a plane wave of amplitude $a^{i}$ and incident angle $\theta^{i}$, is \cite{tr71,wi95a};
\begin{equation}\label{4-010}
u^{[0]}(r,\phi)=u^{i}(r,\phi)+u^{r}(r,\phi)+u^{s}(r,\phi)~,
\end{equation}
wherein:
\begin{equation}\label{4-020}
u^{i}(r,\phi)=a^{i}\exp[ik^{[0]}r\sin(\theta^{i}+\phi)]~~,~~u^{r}(r,\phi)=a^{i}\exp[ik^{[0]}r\sin(\theta^{i}-\phi)]~,
\end{equation}
and by separation-of-variables
\begin{equation}\label{4-030}
u^{s}(r,\phi)=\sum_{m=0}^{\infty}b_{m}H_{m}^{(1)}(k^{[0]}r)\cos~m\phi~;~\forall\mathbf{x}\in\Omega_{0}~,
\end{equation}
\begin{equation}\label{4-040}
u^{[1]}(r,\phi)=\sum_{m=0}^{\infty}a_{m}J_{m}(k^{[1]}r)\cos~m\phi~;~\forall\mathbf{x}\in\Omega_{1}~.
\end{equation}
Note that (\ref{4-030}) previously applied only for $\forall\mathbf{x}\in\Omega_{0}^{+}$ whereas now it applies for $\forall\mathbf{x}\in\Omega_{0}$ because $\Omega_{0}^{+}$=$\Omega_{0}$ when the basin is of semicircular shape. Note also that the expression of the displacement field (\ref{4-040}) is bounded at all points of $\Omega_{1}$ and on $\Gamma_{b}+\Gamma_{m}$, as is implicitly-assumed in the statement of the boundary-value problem.

We recall (\ref{3-090}) that:
\begin{equation}\label{4-050}
B(\phi)=\sum_{m=0}^{\infty}b_{m}\exp(-im\pi/2)\cos(m\phi)~.
\end{equation}
so that:
\begin{equation}\label{4-060}
\mathcal{J}=\int_{\pi}^{2\pi}\|B(\phi)\|^{2}\frac{d\phi}{\pi}=
\sum_{m=0}^{\infty}b^{*}_{m}\sum_{n=0}^{\infty}b_{n}\exp[i(m-n)\pi/2]
\int_{\pi}^{2\pi}\cos(m\phi)\cos(n\phi) d\phi
~.
\end{equation}
But
\begin{equation}\label{4-070}
\int_{\pi}^{2\pi}\cos(m\phi)\cos(n\phi) \frac{d\phi}{\pi}=\frac{\delta_{mn}}{\epsilon_{m}}
~,
\end{equation}
wherein $\delta_{mn}$ is the Kronecker delta symbol, and $\epsilon_{m}$ the Neumann factor, so that
\begin{equation}\label{4-080}
\mathcal{J}=\int_{\pi}^{2\pi}\|B(\phi)\|^{2}\frac{d\phi}{\pi}=
\sum_{m=0}^{\infty}\frac{\|b_{m}\|^{2}}{\epsilon_{m}}
~.
\end{equation}
Also:
\begin{multline}\label{4-090}
\mathcal{K}=2\Re~\big[a^{i*}B(3\pi/2+\theta^{i})\big]=a^{i*}B(3\pi/2+\theta^{i})+a^{i}B^{*}(3\pi/2+\theta^{i})=\\
\sum_{m=0}^{\infty}\Big[a^{i*}b_{m}\exp(-im\pi/2)+a^{i}b_{m}^{*}\exp(im\pi/2)\Big]\cos\big(m(3\pi/2+\theta^{i})\big)=\\
\sum_{m=0}^{\infty}2\Re\Big[\Big(a^{i*}b_{m}\Big)\exp(-im\pi/2)\Big]\cos\big(m(3\pi/2+\theta^{i})\big)
~,
\end{multline}
and:
\begin{multline}\label{4-100}
\mathcal{L}=\frac{\mu^{[1]}}{2\mu^{[0]}}\Im\int_{\Gamma_{b}}u^{[1]*}(\mathbf{x})\boldsymbol{\nu}_{1}\cdot\nabla u^{[1]}(\mathbf{x})d\Gamma=-\frac{\mu^{[1]}}{2\mu^{[0]}}\Im\int_{\pi}^{2\pi}u^{[1]*}(a,\phi)u_{,r}^{[1]*}(a,\phi)ad\phi=\\
-\Im\frac{k^{[1]}\pi a\mu^{[1]}}{2\mu^{[0]}}\sum_{m=0}^{\infty}a^{*}_{m}\big(J_{m}(k^{[1]}a)\big)^{*}
\sum_{n=0}^{\infty}a_{m}^{*}\dot{J}_{n}(k^{[1]}a)\int_{\pi}^{2\pi}\cos(m\phi)\cos(n\phi)\frac{d\phi}{\pi}=\\
-\frac{\pi k^{[0]}a}{2\mu^{[0]}k^{[0]}}
\sum_{m=0}^{\infty}\frac{\|a_{m}\|^{2}}{\epsilon_{m}}\Im\big[\mu^{[1]}k ^{[1]}J_{m}^{*}(k^{[1]}a)\dot{J}_{m}(k^{[1]}a)\Big]
~,
\end{multline}
wherein $\dot{J}_{m}(\zeta)=\big(J_{m}(\zeta)\big)_{,\zeta}$ and use has been made of (\ref{4-070}).

We cannot go beyond this point without making use of the explicit, rigorous solutions for $a_{m}$ and $b_{m}$. These are \cite{tr71,wi95a}:
\begin{equation}\label{4-110}
a_{m}=\frac{N^{a}_{m}}{D_{m}}~~,~~b_{m}=\frac{N^{b}_{m}}{D_{m}}~,
\end{equation}
\begin{equation}\label{4-110}
N^{b}_{m}=
-2a^{i}\epsilon_{m}
\exp(-im\pi/2)\cos(m\phi^{i})
\left[-\mu^{[1]}k^{[1]}\dot{J}_{m}(k^{[1]}a)J_{m}(k^{[0]}a)+
\mu^{[0]}k^{[0]}J_{m}(k^{[1]}a)\dot{J}_{m}(k^{[0]}a)\right]
~,
\end{equation}
\begin{equation}\label{4-120}
N^{a}_{m}=
-2a^{i}\epsilon_{m}
\exp(-im\pi/2)\cos(m\phi^{i})
\left[\mu^{[0]}k^{[0]}\dot{J}_{m}(k^{[0]}a)H_{m}^{(1)}(k^{[0]}a)-
\mu^{[0]}k^{[0]}J_{m}(k^{[0]}a)\dot{H}_{m}^{(1)}(k^{[0]}a)\right]
~,
\end{equation}
or, with the help of (9.1.17) in \cite{as68}:
\begin{equation}\label{4-130}
N^{a}_{m}=
-2a^{i}\epsilon_{m}
\exp(-im\pi/2)\cos(m\phi^{i})
\left[\mu^{[0]}k^{[0]}\frac{-2i}{\pi k^{[0]}a}\right]
~,
\end{equation}

in which
\begin{equation}\label{4-140}
\phi^{i}=\frac{3\pi}{2}-\theta^{i}
\end{equation}
and
\begin{equation}\label{4-150}
D_{m}=
\mu^{[0]}k^{[0]}J_{m}(k^{[1]}a)\dot{H}_{m}^{(1)}(k^{[0]}a)-
\mu^{[1]}k^{[1]}\dot{J}_{m}(k^{[1]}a)H_{m}^{(1)}(k^{[0]}a)
~.
\end{equation}
It follows that:
\begin{equation}\label{4-160}
\mathcal{J}=\sum_{m=0}^{\infty}
\frac{\|N^{b}_{m}\|^{2}}
{\epsilon_{m}\|D_{m}\|^{2}}
~,
\end{equation}
\begin{equation}\label{4-170}
\mathcal{K}=
2\Re\sum_{m=0}^{\infty}a^{i*}\frac{N^{b}_{m}}{D_{m}}\exp(im\pi/2)\cos(m\phi^{i})
~,
\end{equation}
\begin{equation}\label{4-180}
\mathcal{L}=\Big(\frac{-\pi k^{[0]}a}{2z^{[0]}}\Big)
\sum_{m=0}^{\infty}\frac{\|N^{a}_{m}\|^{2}}{\epsilon_{m}\|D_{m}\|^{2}}
\Im\big[z^{[1]}J^{*}(1)\dot{J}(1)\big]
~,
\end{equation}
wherein we adopt henceforth the short-hand notations: $J(l)=J_{m}(k^{[l]}a)$, $H(l)=H_{m}^{(1)}(k^{[l]}a)$ and $z^{[l]}=\mu^{l]}k{l]}$. The conservation of flux law can  now be expressed as
\begin{equation}\label{4-190}
-\mathcal{J}+\mathcal{L}=\mathcal{K}
~.
\end{equation}
Let us now go into more detail. On account of (\ref{4-130}) we have
\begin{equation}\label{4-200}
\|N^{a}_{m}\|^{2}=
-8\|a^{i}\|^{2}\epsilon_{m}
\left[\frac{z^{[0]}}{\pi k^{[0]}a}\right]^{2}
\cos^{2}(m\phi^{i})
~,
\end{equation}
so that (\ref{4-180}) becomes
\begin{equation}\label{4-210}
\mathcal{L}=-\frac{2z^{[0]}}{\pi k^{[0]}}\sum_{m=0}^{\infty}\Big(\frac{4\epsilon_{m}\|a^{i}\|^{2}\cos^{2}(m\phi^{i})}{\|D_{m}\|^{2}}\Big)
\Im\big[z^{[1]}J^{*}(1)\dot{J}(1)\big]
~,
\end{equation}
from which we again find, but more overtly, that $\mathcal{L}=0$ when $\Im(k^{[1]})=0$, which occurs when the medium in the basin is non-lossy, and entails $-\mathcal{J}=\mathcal{K}$.

Eq. (\ref{4-160}) leads to:
\begin{equation}\label{4-220}
\mathcal{J}=\sum_{m=0}^{\infty}\Big(\frac{4\epsilon_{m}\|a^{i}\|^{2}\cos^{2}(m\phi^{i})}{\|D_{m}\|^{2}}\Big)
\|-z^{[1]}J(0)\dot{J}(1)-z^{[0]}J(1)\dot{J}(0)\|^{2}
~,
\end{equation}
or
\begin{multline}\label{4-230}
\mathcal{J}=\sum_{m=0}^{\infty}\Big(\frac{4\epsilon_{m}\|a^{i}\|^{2}\cos^{2}(m\phi^{i})}{\|D_{m}\|^{2}}\Big)\times\\
\Big[\|z^{[1]}\|^{2}J^{2}(0)\|\dot{J}(1)\|^{2}-2z^{[0]}J(0)\dot{J}(0)\Re\big(z^{[1]}\dot{J}(1)J^{*}(1)\big)+
(z^{[0]})^{2})\dot{J}^{2}(0)\|J(1)\|^{2}\Big]~.
\end{multline}
Eq. (\ref{4-170}) gives:
\begin{multline}\label{4-240}
\mathcal{K}=-\frac{1}{2}\sum_{m=0}^{\infty}\Big(\frac{4\epsilon_{m}\|a^{i}\|^{2}\cos^{2}(m\phi^{i})}{\|D_{m}\|^{2}}\Big)\times\\
\Big\{
\big[-z^{[1]}J(0)\dot{J}(1)+z^{[0]}J(1)\dot{J}(0)\big]D_{m}^{*}+
\big[-z^{[1]*}J(0)\dot{J}^{*}(1)+z^{[0]}J^{*}(1)\dot{J}(0)\big]D_{m}\big]
\Big\}~,
\end{multline}
On account of the fact that $H_{m}(l)=J _{m}(l)+iY_{m}(l)$ (or $H(l)=J (l)+iY(l)$), with $Y_{m}(l)$ the Neumann function \cite{as68}, we further find:
\begin{multline}\label{4-210}
\mathcal{K}=-\frac{1}{2}\sum_{m=0}^{\infty}\Big(\frac{4\epsilon_{m}\|a^{i}\|^{2}\cos^{2}(m\phi^{i})}{\|D_{m}\|^{2}}\Big)\times\\
\Big\{-z^{[0]}2\Re\big(z^{[1]}\dot{J}(1)J^{*}(1)\dot{H}^{*}(0)\big)+\|z^{[1]}\dot{J}(1)\|^{2} 2 J(0)\Re \big(H^{*}(0)\big)+\\
(z^{[0]})^{2}\|J(1)\|^{2} \dot{J}(0)2\Re\big(\dot{H}(0)\big)-z(0)\dot{J}(0)2\Re\big(z(1)\dot{J}(1)J^{*}(1)H(0)\big)
\Big\}=
-\sum_{m=0}^{\infty}\Big(\frac{4\epsilon_{m}\|a^{i}\|^{2}\cos^{2}(m\phi^{i})}{\|D_{m}\|^{2}}\Big)\times\\
\Big\{\|z^{[1]}\|^{2}J^{2}(0)\|\dot{J}(1)\|^{2}-
2z^{[0]}J(0)\dot{J}(0)\Re\big[z^{[1]}\dot{J}(1)J^{*}(1)\big]+(z^{[0]})^{2}\dot{J}^{2}(0)\|J(1)\|^{2}
\Big\}-\\
\sum_{m=0}^{\infty}\Big(\frac{4\epsilon_{m}\|a^{i}\|^{2}\cos^{2}(m\phi^{i})}{\|D_{m}\|^{2}}\Big)
\Big\{z^{[0]}\Big[J(0)\dot{Y}(0)-\dot{J}(0)Y(0)\Big]\Im\Big[z^{[1]}\dot{J}(1)J^{*}(1)\Big]
\Big\}=\\
-\mathcal{J}-\frac{2z^{[0]}}{\pi k^{[0]}a}\sum_{m=0}^{\infty}\Big(\frac{4\epsilon_{m}\|a^{i}\|^{2}\cos^{2}(m\phi^{i})}{\|D_{m}\|^{2}}\Big)\Im\Big[z^{[1]}\dot{J}(1)J^{*}(1)\Big]=
-\mathcal{J}+\mathcal{L}
\end{multline}
Thus, the exact solutions in the case of the semi-circular basin, satisfy the  conservation of flux relation $\mathcal{K}=-\mathcal{J}+\mathcal{L}$ whether the basin is lossy (in which case $\mathcal{L}\ne 0$) or non lossy (in which case $\mathcal{L}= 0$).
%%%%%%%%%%%%%%%%%%%%%%%%%%%%%%%%%%%%%%%%%%%%%%%
\section{Conclusion}
We have thus established a conservation law concerning the response of a sedimentary basin of arbitrary shape to a plane seismic wave and shown that the well-known, exact solutions for the sedimentary basin of semi-circular shape are such as to  satisfy the equation which translates this law. Moreover, we have shown that the three terms in this law can be interpreted physically as being normalized scattered flux, normalized absorbed flux and normalized input flux. This law shows that: (1) when the basin is empty (i.e., is a canyon), the scattered flux equals the input flux, (2) when the basin is filled with a non-lossy medium, the scattered flux again equals the input flux, and (3) when the basin is filled with a lossy medium, the scattered flux plus the absorbed flux equals the input flux.

Any solution, be it theoretical or numerical, of such a scattering problem, for a totally filled or totally empty basin of arbitrary shape, must satisfy this conservation law. Thus, the more such a solution violates this law, the less valid is the said solution. But, one should be aware of the fact that a faulty solution can satisfy the conservation of flux law, so that the latter is a necessary, but not sufficient, condition for a solution to be satisfactory.
%%%%%%%%%%%%%%%%%%%%%%%%%%%%%%%%%%%%%%%%%%%%%%%%%%%%%%%%%%%%%%%%%%%%%%%%%%%%%%%%%%%%%%%%%%%%%%%%%%%%%%%%%%%
%\clearpage
%\newpage
%%%%%%%%%%%%%%%%%%%%%%%%%%%%%%%%%%%%%%%%%%%%%%%%%%%%%%%%%%%%%%%%%%%%%%%%%%%%%%%%%%%%%%%%%%%%%%%%%%%%%%%%%%%%

%%%%%%%%%%%%%%%%%%%%%%%%%%%%%%%%%%%%%%%%%%%%%%%%%%%%%%%%%%%%%%%%%%%%%%%%%%%%%%%%%%%%%%%%%%%%%%%%%%%%%%%%%%%%%%%%%%%%%%%%%%%%%%%%%%
\end{document}